\documentclass[proceedings,conferenceproceedings,submit,moreauthors,pdftex,10pt,a4paper]{mdpi} 

\usepackage{amssymb}
\def\be{\begin{equation}}
\def\ee{\end{equation}}
\def\ba{\begin{eqnarray}}
\def\ea{\end{eqnarray}}

\firstpage{1} 
\pubvolume{xx}
\issuenum{1}
\articlenumber{5}
\pubyear{2018}
\copyrightyear{2018}
\history{Received: date; Accepted: date; Published: date}

\Title{Anisotropic hydrodynamics for Au-Au collisions at 
200 GeV$^{\dagger}$}

\Author{Mubarak Alqahtani $^{1,}$*, Dekrayat Almaalol $^{2}$ and Michael Strickland $^{2}$}

\AuthorNames{Mubarak Alqahtani , Dekrayat Almaalol  and Michael Strickland}

\address{%
$^{1}$ \quad Department of Basic Sciences, College of Education, Imam Abdulrahman Bin Faisal University, Dammam 34212, Saudi Arabia.\\
$^{2}$ \quad Department of Physics, Kent State University, Kent, OH 44242 United States}

\corres{Correspondence: maqahtani@iau.edu.sa}

\firstnote{Presented at Hot Quarks 2018 - Workshop for young scientists on the physics of ultrarelativistic nucleus-nucleus collisions, Texel, The Netherlands, September 7-14 2018}

\abstract{In this proceedings, we review the basics of quasiparticle anisotropic hydrodynamics (aHydroQP). Then we present phenomenological comparisons between 3+1d quasiparticle anisotropic hydrodynamics and experimental data from RHIC experiments at 200 GeV Au-Au collisions. We show that 3+1d aHydroQP model is able to describe the experimental results quite well for the spectra, multiplicity, and elliptic flow for charged particles in different centrality classes.  }


\begin{document}
\section{Introduction}
Heavy-ion collision experiments at the Relativistic Heavy Ion Collider (RHIC) and Large Hadron Collider (LHC) create and study the quark-gluon plasma (QGP). In the early years after confirming the existence of QGP, ideal hydrodynamics was used to describe the collective behavior seen in these experiments  \cite{Huovinen:2001cy}. Later on, viscous hydrodynamics was used to  take into account the dissipative effects \cite{Romatschke:2007mq,Ryu:2015vwa,Niemi:2011ix}. However, the QGP is a highly momentum anisotropic plasma at early times after the impact which motivates introducing anisotropic hydrodynamics \cite{Florkowski:2010cf,Martinez:2010sc,Martinez:2012tu,Ryblewski:2012rr,Bazow:2013ifa,Nopoush:2014pfa,Nopoush:2014qba}. Recently, the 3+1d quasiparticle anisotropic hydrodynamics model was introduced and compared to experimental data at different energies \cite{Alqahtani:2017jwl,Alqahtani:2017tnq,Almaalol:2018gjh}. For more details about anisotropic hydrodynamics, we refer the reader to \cite{Strickland:2014pga,Alqahtani:2017mhy}.

In this proceedings contribution, we will first introduce 3+1d quasiparticle anisotropic hydrodynamics (aHydroQP) \cite{Alqahtani:2015qja,Alqahtani:2016rth}. Then, we will present some phenomenological comparisons between 3+1d aHydroQP and some heavy-ion observables for Au-Au collisions at 200 GeV from different RHIC experiments. We show the spectra, multiplicity, and elliptic flow for charged particles for which aHydroQP shows good agreement with data \cite{Almaalol:2018gjh}. 

\section{3+1d quasiparticle anisotropic hydrodynamics}
In anisotropic hydrodynamics, the one-particle distribution function is assumed to be of the form
\be
f(x,p) = f_{\rm iso}\!\left(\frac{1}{\lambda}\sqrt{p_\mu \Xi^{\mu\nu} p_\nu}\right) ,
\label{eq:genf}
\ee
where $ \lambda $ can be identified with the temperature in the isotropic equilibrium limit and $\Xi^{\mu\nu}$ is the anisotropy tensor \cite{Nopoush:2014pfa}
\be
\Xi^{\mu\nu} = u^\mu u^\nu + \xi^{\mu\nu} - \Delta^{\mu\nu} \Phi \, ,
\ee
Here,  $u^\mu$ is the fluid four-velocity, $\xi^{\mu\nu}$ is a symmetric and traceless anisotropy tensor, and $\Phi$ is the degree of freedom associated with the bulk pressure \cite{Alqahtani:2017mhy}.

Ignoring the off-diagonal components, in the local rest frame the distribution function in Eq.~(\ref{eq:genf}) can be written compactly in the following form 
\be
f(x,p) =  f_{\rm eq}\!\left(\frac{1}{\lambda}\sqrt{\sum_i \frac{p_i^2}{\alpha_i^2} + m^2}\right) ,
\label{eq:fform}
\ee
where $i\in \{x,y,z\}$ and $\alpha_i \equiv (1 + \xi_i + \Phi)^{-1/2} \,$. We note that by taking $\alpha_x=\alpha_y=\alpha_z=1$ and $\lambda=T$ one recovers the isotropic equilibrium distribution function. We also note that $m(T)$ is a single effective mass which is a function of temperature and tuned to match the equation of state of QCD.

To obtain, the dynamical equations for  3+1d  aHydroQP one can take moments of the Boltzmann equation \cite{Alqahtani:2015qja}.
\be
p^\mu \partial_\mu f(x,p)+\frac{1}{2}\partial_i m^2\partial^i_{(p)} f(x,p)=-C[f(x,p)]\,,
\label{eq:boltzmanneq}
\ee
where $C[f(x,p)]$ is the collisional kernel which is taken to be in the relaxation time approximation \cite{Alqahtani:2015qja}. 
\begin{figure*}[t!]
\centerline{
\includegraphics[width=0.99\linewidth]{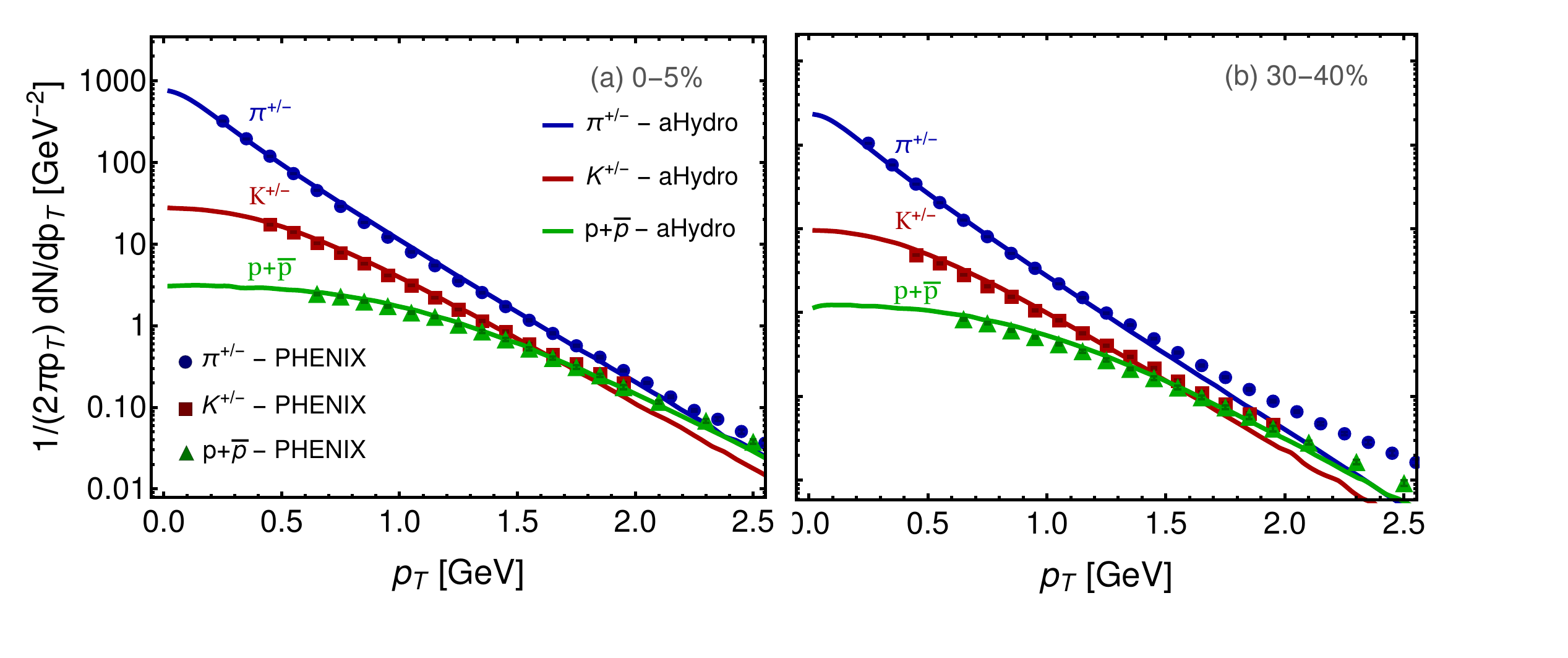}
}
\caption{Pion, kaon, and proton spectra predicted by aHydroQP compared to experimental observations by the PHENIX collaboration \cite{Adler:2003cb}.  The panels show the centrality classes (a) 0-5\% and (b) 30-40\%. 
}
\label{fig:spectra}
\end{figure*}
\section{Phenomenological results}
\label{}

Next, we turn to presenting comparisons between 3+1d quasiparticle anisotropic hydrodynamics and  experimental data from  200 GeV Au-Au collisions.  In Ref.~\cite{Almaalol:2018gjh}, we presented a number of observables: spectra, identified particle multiplicities as a function of centrality, multiplicity, elliptic flow for charged and identified particles, and the pseudorapidity dependence of the integrated elliptic flow. In this proceedings, we will present only the spectra, the multiplicity, and the elliptic flow for charged particles due to the space limitation. For more centrality classes or other observables, we refer the reader to Ref.~\cite{Almaalol:2018gjh}.

In Fig.~\ref{fig:spectra}, we show the spectra of pions, kaons, and protons as a function of the transverse momentum $p_T$. In the left panel, we show the 0-5\% centrality class while in the right panel we show the 30-40\% centrality class. From both panels, we see that aHydroQP agrees with the data quite well.

\begin{figure*}[t!]
\centerline{
\includegraphics[width=.48\linewidth]{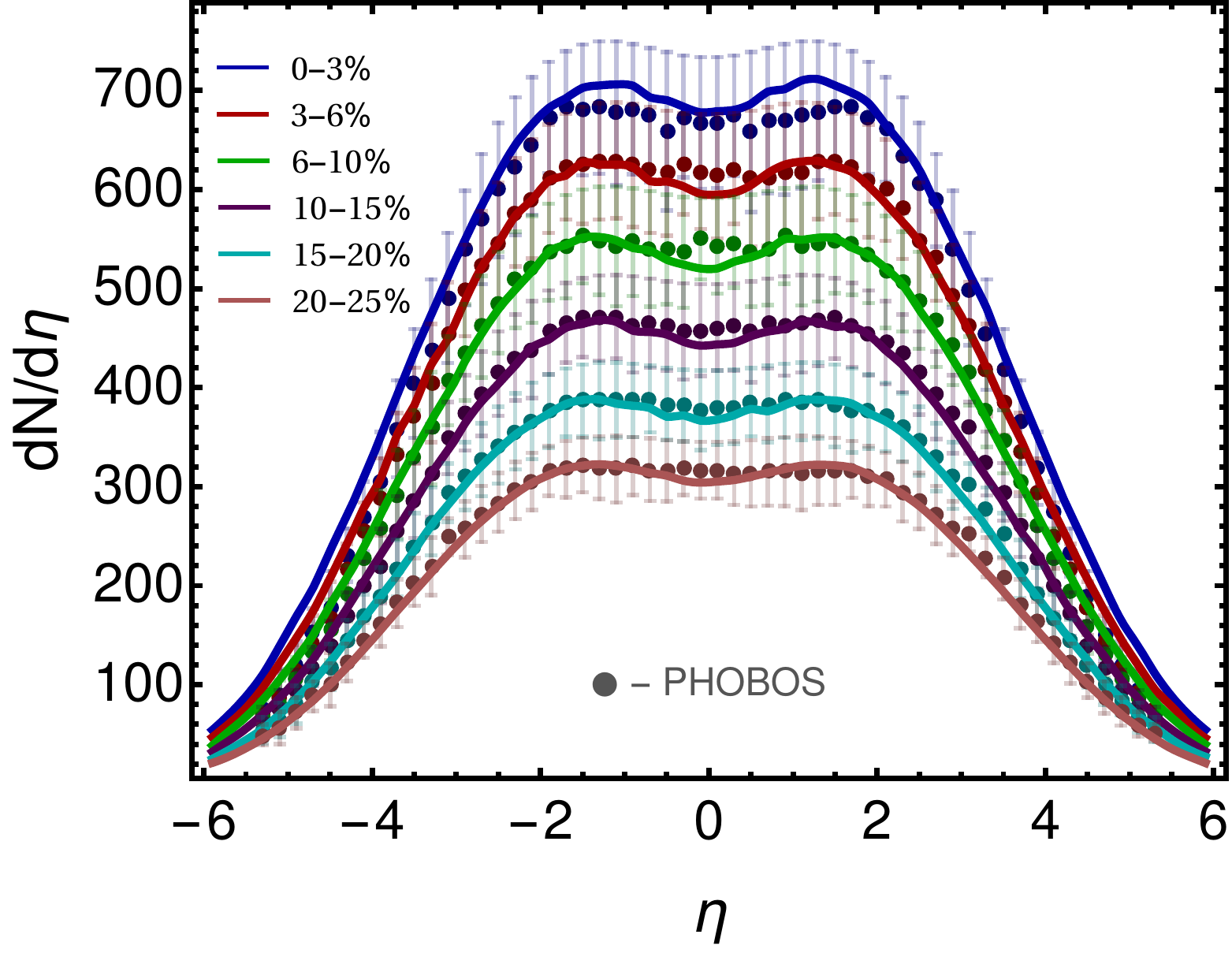}
\includegraphics[width=0.48\linewidth]{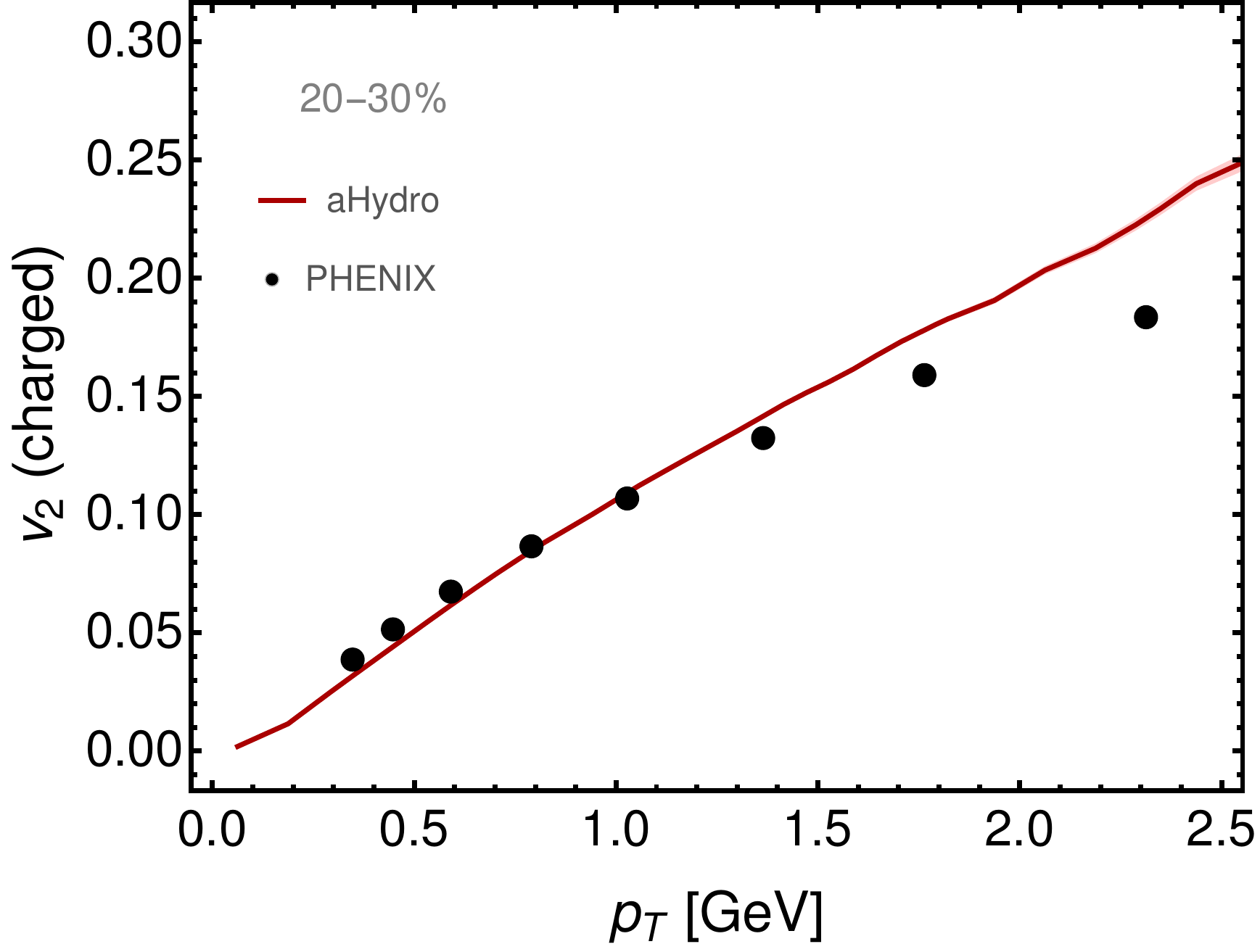}
}
\caption{In panel (a), a comparison of the charged particle multiplicity in different centrality classes (0-25\%) is shown between aHydroQP and experimental data which is taken from the PHOBOS collaboration \cite{Alver:2010ck}. In panel (b), the elliptic flow for charged particles in 20-30\% centrality class is shown where 3+1d aHydroQP predictions is compared to data taken from the PHENIX collaboration \cite{Adare:2006ti}. Figure is taken from \cite{Almaalol:2018gjh}.}
\label{fig:multiplicity_v2}
\end{figure*}

Next, in Fig.~\ref{fig:multiplicity_v2}-a, we show comparisons of the charged particle multiplicity as a function of pseudorapidity predicted by 3+1d aHydroQP and experimental data. As can be seen from the figure, in a wide range of centrality classes, the agreement between aHydroQP and the experimental results is quite good. In Fig.~\ref{fig:multiplicity_v2}-b, we present the elliptic flow for charged particles as a function of transverse momentum in the 20-30\% centrality class. We find that our model shows good agreement with the experimental results. The data in the left and  right panels are from the PHOBOS collaboration \cite{Alver:2010ck} and  the PHENIX collaboration \cite{Adare:2006ti}, respectively.

The extracted fitting parameters that were used in the above phenomenological comparisons are $T_0 (\tau_0=0.25$\,fm/c$)= 455$ MeV, $\eta/s = 0.179$, and the freeze-out temperature used was  \mbox{$T_{\rm FO} = 130$ MeV}.

\section{Conclusions}
In this proceedings contribution, we presented phenomenological comparisons between 3+1d quasiparticle anisotropic hydrodynamics and Au-Au collisions at 200 GeV. We showed the spectra, multiplicity, and elliptic flow for charged particle in some centrality classes. We demonstrated that aHydroQP agrees with the data quite well for many observables. Finally, we listed the extracted fitting parameters that were used in these comparisons.


\vspace{6pt} 




\acknowledgments{M.~Alqahtani was supported by Imam Abdulrahman Bin Faisal University, Saudi Arabia.  D. Almaalol was supported by a fellowship from the University of Zawia, Libya. M.~Strickland was supported by the U.S. Department of Energy, Office of Science, Office of Nuclear Physics under Award No. DE-SC0013470.}
\authorcontributions{M. Alqahtani , D. Almaalol, and M.~Strickland  contributed to writing the code and the writing of the original paper and to this proceedings.}

\conflictsofinterest{The authors declare no conflict of interest.} 




\reftitle{References}


\bibliographystyle{mdpi}
\externalbibliography{yes}
\bibliography{HQ18}



\end{document}